\documentclass[reprint,aps,prd,onecolumn,notitlepage,nofootinbib,showpacs,preprintnumbers,superscriptaddress]{revtex4-1}
\usepackage[T1]{fontenc}
\usepackage[utf8]{inputenc}
\usepackage{bm}
\usepackage{amsmath}
\usepackage{esint}
\usepackage{color}
\usepackage{graphicx}
\PassOptionsToPackage{normalem}{ulem}
\usepackage{ulem}

\usepackage{hyperref}

\begin{document}
	
	\title{Spatially covariant gravity: Perturbative analysis and field transformations}
	
	\author{Xian Gao}%
	\email[Email: ]{gaoxian@mail.sysu.edu.cn}
	\affiliation{%
		School of Physics and Astronomy, Sun Yat-sen University, Guangzhou 510275, China}
	
	\author{Chao Kang}%
	\email[Email: ]{kangch@mail2.sysu.edu.cn}
	\affiliation{%
		School of Physics and Astronomy, Sun Yat-sen University, Guangzhou 510275, China}
	
	\author{Zhi-Bang Yao}%
	\email[Email: ]{yaozhb@mail2.sysu.edu.cn}
	\affiliation{%
		School of Physics and Astronomy, Sun Yat-sen University, Guangzhou 510275, China}
	
	\date{February 20, 2019}
	
	\begin{abstract}
		We make a perturbative analysis of the number of degrees of freedom in a large class of metric theories respecting spatial symmetries, of which the Lagrangian includes kinetic terms of both the spatial metric and the lapse function. We show that, as long as the kinetic terms are degenerate, the theory propagates a single scalar mode at the linear order in perturbations around a Friedmann-Robertson-Walker background. Nevertheless, an unwanted mode will reappear pathologically, either at nonlinear orders around the Friedmann-Robertson-Walker background, or at linear order around an inhomogeneous background. In both cases, it turns out that a consistency condition has to be imposed in order to remove the unwanted mode. This perturbative approach provides an alternative and also complementary point of view of the conditions derived in a Hamiltonian analysis. We also discuss the relation under field redefinitions, between theories with and without the time derivative of the lapse function.
	\end{abstract}
	
	\maketitle

\section{Introduction}

The scope of scalar-tensor theories has been extended significantly in the past decade.
In particular, higher derivatives of the scalar field(s) as well as novel couplings between scalar field(s) and the gravity have attracted much attention.
The representative achievements are the $k$-essence \cite{ArmendarizPicon:1999rj,Chiba:1999ka} and the Horndeski/Galileon theory \cite{Horndeski:1974wa,Deffayet:2011gz,Kobayashi:2011nu}, as well as more general higher-order derivative theories with degeneracies  \cite{Gleyzes:2014dya,Gleyzes:2014qga,Langlois:2015cwa,Langlois:2015skt,Crisostomi:2016tcp,Crisostomi:2016czh,Achour:2016rkg,Ezquiaga:2016nqo,Motohashi:2014opa,Motohashi:2016ftl,Deffayet:2015qwa,deRham:2016wji,BenAchour:2016fzp,Chagoya:2016inc,Crisostomi:2017aim}, which guarantee the absence of the Ostrogradsky ghost in the presence of higher derivatives \cite{Ostrogradsky:1850fid,Woodard:2015zca}.
See Refs. \cite{Langlois:2018dxi,Kobayashi:2019hrl,Quiros:2019ktw,Amendola:2019laa} for recent reviews on this progress.

An alternative approach to extending the scalar-tensor theories is to construct theories that do not respect the full symmetry of General Relativity (GR).
In many situations, the ``scalar'' is merely an effective scalar-type degree of freedom instead of a covariant scalar field.
This idea can be traced back to the  effective field theory of inflation \cite{Creminelli:2006xe,Cheung:2007st} and of dark energy \cite{Creminelli:2008wc,Gubitosi:2012hu,Bloomfield:2012ff,Gleyzes:2013ooa,Bloomfield:2013efa,Gleyzes:2014rba,Gleyzes:2015pma,Gleyzes:2015rua}, as well as Ho\v{r}ava gravity \cite{Horava:2009uw,Blas:2009qj}. 
We may refer to such theories as spatially covariant gravity theories since they are metric theories that respect only spatial symmetries instead of the full spacetime symmetries.

If we take the spatial covariance as our starting point, it is natural to extend the scope of scalar-tensor theories by exploring spatially covariant theories of gravity as general as we can. 
In Ref. \cite{Gao:2014soa}, a general framework for spatially covariant gravity theories with the Lagrangian composed of polynomials of the extrinsic curvature $K_{ij}$ was proposed.
Through a Hamiltonian analysis in a perturbative manner \cite{Gao:2014fra} and in a nonperturbative manner \cite{Saitou:2016lvb} (see also Ref. \cite{Lin:2014jga} for a related analysis), it has been shown that there are at most three physical degrees of freedom.
Such theories go far beyond previously known scalar-tensor theories, at least in the unitary gauge with $t = \phi(t,\vec{x})$, and also lead to novel features in cosmological applications \cite{Kobayashi:2015gga,Fujita:2015ymn,Yajima:2015xva,Akita:2015mho} (see also Refs. \cite{Cai:2016thi,Cai:2017dxl,He:2018mse,Ye:2019sth} for related studies).
This framework was also generalized in Ref. \cite{Gao:2018izs}, which introduced an additional nondynamical scalar field without changing the number of degrees of freedom.

A further extension of the framework was made in Ref. \cite{Gao:2018znj}, by including the time derivative of the lapse function $N$.
As has been illustrated in Ref. \cite{Gao:2018znj}, both velocities of the spatial metric $h_{ij}$ and the lapse function $N$ are natural geometric quantities from the geometric point of view.
In fact, time derivative of the lapse function arises in higher-order derivative theories with degeneracies \cite{Langlois:2017mxy}.
On the other hand, higher-order derivative scalar-tensor theories without the Ostrogradsky ghosts \cite{Bettoni:2013diz,Zumalacarregui:2013pma,Domenech:2015tca,Crisostomi:2016tcp,Achour:2016rkg,Arroja:2015wpa,Takahashi:2017pje} can be generated from field transformations, such as the disformal transformation \cite{Bekenstein:1992pj} or mimetic transformation \cite{Chamseddine:2013kea} (see Ref. \cite{Sebastiani:2016ras} for a review on mimetic gravity).
Typically, the transformed theories acquire a time derivative of the lapse function.

Contrary to the theories in Ref. \cite{Gao:2014soa}, generally including $\dot{N}$ in the Lagrangian will introduce an extra scalar-type degree of freedom.
Nevertheless, according to a Hamiltonian analysis in Ref. \cite{Gao:2018znj}, two conditions are found in order to get rid of such an unwanted scalar mode.
One condition requires the kinetic terms of $h_{ij}$ and $N$ to be degenerated, which implies the existence of a primary constraint in the language of Hamiltonian constraint analysis.
However, due to the loss of general covariance, the primary constraint is not necessarily associated with a secondary constraint.
As a result, a second condition, which we dub the consistency condition, must be imposed in order to ensure the existence of an additional secondary constraint and to fully removed the unwanted mode.

Although counting the number of degrees of freedom is well performed through the Hamiltonian analysis, the explanation is quite formal (if not obscure), especially for those who are not familiar with the terminology of constraint analysis.
The purpose of this paper is thus to provide an alternative derivation and also a complementary understanding of such two conditions in order to evade the unwanted mode.
The idea is to examine how such conditions arise in a perturbative manner, at the level of the Lagrangian.
First, starting from an unconstrained Lagrangian, we consider linear perturbations around a Friedmann-Robertson-Walker (FRW) background in order to check under which condition(s) there is only a single scalar mode that propagates.
It is possible that a mode that disappears in the lower order in perturbation theory reappears in higher orders or in a nontrivial background.
To check if this is the case and to check if additional condition(s) should be imposed, we discuss two situations.
One is the second-order perturbation around a FRW background, the other is the linear perturbation around an inhomogeneous background.

The paper is organized as following.
In the next section, we make a brief review of the two conditions derived in Ref. \cite{Gao:2018znj}.
In Sec.\ref{sec:pert}, we consider linear perturbations around the FRW background based on a simple prototype action, and show that the degeneracy condition arises in order to evade the unwanted mode at linear order.
In Sec.\ref{sec:cons}, we show that the unwanted mode will reappear either at second order around the FRW background or at linear order but around an inhomogeneous background.
In this case, an additional condition which is exactly the consistency condition, naturally arises in order to evade the unwanted mode.
In Sec.\ref{sec:quad_trans}, we construct a more general action that is quadratic in time derivatives of $h_{ij}$ and $N$, and also show that such a specific model can be transformed from a model without the time derivative of $N$ through a field transformation.
Section \ref{sec:con} concludes.

\section{Spatially Covariant Gravity with 3 degrees of freedom}

A wide class of spatially covariant theories of gravity was considered in Ref. \cite{Gao:2018znj}, of which the action takes the form
	\begin{equation}
	S =\int\mathrm{d}t\mathrm{d}^{3}x \, N\sqrt{h}\mathcal{L}\left(t,N,h_{ij},F,K_{ij},R_{ij},\nabla_{i}\right),\label{S_scg}
	\end{equation}
where $N$ and $h_{ij}$ are the lapse function and spatial metric in the Arnowitt-Deser-Misner (ADM) formalism, 
	\begin{eqnarray}
	F & := & \pounds_{\bm{n}}N=\frac{1}{N}\left(\dot{N}-\pounds_{\vec{N}}N\right), \label{F_def}\\
	K_{ij} & := & \frac{1}{2}\pounds_{\bm{n}}h_{ij}=\frac{1}{2N}\left(\dot{h}_{ij}-\pounds_{\vec{N}}h_{ij}\right), \label{Kij_def}
	\end{eqnarray}
where an overdot denotes a derivative with respect to the physical time $t$. In (\ref{S_scg}), the Lagrangian is a general function of $t$, the lapse function $N$, the spatial metric $h_{ij}$, and their ``velocities'' $F$ and $K_{ij}$ defined in (\ref{F_def}) and (\ref{Kij_def}), as well as the spatial curvature $R_{ij}$. In (\ref{F_def}) and (\ref{Kij_def}), $\pounds_{\bm{n}}$ and $\pounds_{\vec{N}}$ denote the projected Lie derivatives on the spatial hypersurfaces, with respect to the normal vector of the spatial hypersurfaces $n^{\mu}$ and to the shift vector $N^{\mu}$, respectively.

Generally the action (\ref{S_scg}) propagates 4 dynamical degrees of freedom: two of the tensor type and another two of the scalar type. The tensor-type degrees of freedom are the transverse-traceless modes of the spatial metric, which are the only propagating modes in GR. 
One of the two scalar modes can be identified to be the longitudinal mode of the spatial metric, and the other scalar mode arises due to the fact that the lapse function $N$ becomes dynamical in general.

In Ref. \cite{Gao:2018znj}, two conditions on the functional structure of the Lagrangian in (\ref{S_scg}) are derived in order to guarantee that there are at most three dynamical modes. In particular, there is only one scalar mode propagating.
One condition requires that the kinetic terms of $N$ and $h_{ij}$ are degenerate, which can be written as
	\begin{equation}
	0\equiv \mathcal{D}\left(\vec{x},\vec{y}\right):=\frac{\delta^{2}S}{\delta F\left(\vec{x}\right)\delta F\left(\vec{y}\right)}-\int\!\mathrm{d}^{3}\vec{x}'\int\!\mathrm{d}^{3}y'\,\frac{\delta^{2}S}{\delta F\left(\vec{x}\right)\delta K_{ij}\left(\vec{x}'\right)}\mathcal{G}_{ij,kl}\left(\vec{x}',\vec{y}'\right)\frac{\delta^{2}S}{\delta K_{kl}\left(\vec{y}'\right)\delta F\left(\vec{y}\right)}, \label{degen_con}
	\end{equation}
where $\mathcal{G}_{ij,kl}(\vec{x},\vec{y})$ is the inverse of the second-order functional derivative of the action $S$ with respect to $K_{ij}$, defined by
	\begin{equation}
	\int\!\mathrm{d}^{3}z\,\mathcal{G}_{mn,ij}\left(\vec{x},\vec{z}\right)\frac{\delta^{2}S}{\delta K_{ij}\left(\vec{z}\right)\delta K_{kl}\left(\vec{y}\right)}\equiv \mathbf{I}_{mn}^{kl}\delta^{3}\left(\vec{x}-\vec{y}\right), \label{calG_inv}
	\end{equation}
with $\mathbf{I}_{kl}^{ij}\equiv \frac{1}{2}\left(\delta_{k}^{i}\delta_{l}^{j}+\delta_{l}^{i}\delta_{k}^{j}\right)$ being the identity in the linear space of $3\times3$
symmetric matrices.
We refer to (\ref{degen_con}) as the degeneracy condition.
The other condition, which we dub the consistency condition, is more involved and given by
	\begin{equation}
	\mathcal{F}\left(\vec{x},\vec{y}\right)=0, \label{consis_con}
	\end{equation}
where 
	\begin{eqnarray}
	\mathcal{F}\left(\vec{x},\vec{y}\right) & := & \frac{1}{N\left(\vec{y}\right)}\frac{\delta^{2}S}{\delta N\left(\vec{x}\right)\delta F\left(\vec{y}\right)}\nonumber \\
	&  & +\frac{1}{N\left(\vec{x}\right)N\left(\vec{y}\right)}\frac{\delta S}{\delta K_{ij}\left(\vec{x}\right)}\int\mathrm{d}^{3}z\,\mathcal{G}_{ij,kl}\left(\vec{x},\vec{z}\right)\frac{\delta^{2}S}{\delta K_{kl}\left(\vec{z}\right)\delta F\left(\vec{y}\right)}\nonumber \\
	&  & -\frac{1}{N\left(\vec{y}\right)}\int\!\mathrm{d}^{3}z\int\mathrm{d}^{3}z'\,\frac{\delta^{2}S}{\delta N\left(\vec{x}\right)\delta K_{ij}\left(\vec{z}\right)}\mathcal{G}_{ij,kl}\left(\vec{z},\vec{z}'\right)\frac{\delta^{2}S}{\delta K_{kl}\left(\vec{z}'\right)\delta F\left(\vec{y}\right)}\nonumber \\
	&  & +\frac{1}{N\left(\vec{x}\right)N\left(\vec{y}\right)}\int\!\mathrm{d}^{3}z\int\!\mathrm{d}^{3}z'\,\frac{\delta^{2}S}{\delta F\left(\vec{x}\right)\delta h_{ij}\left(\vec{z}\right)}2N\left(\vec{z}\right)\mathcal{G}_{ij,kl}\left(\vec{z},\vec{z}'\right)\frac{\delta^{2}S}{\delta K_{kl}\left(\vec{z}'\right)\delta F\left(\vec{y}\right)}\nonumber \\
	&  & +\frac{1}{N\left(\vec{x}\right)N\left(\vec{y}\right)}\int\!\mathrm{d}^{3}x'\int\!\mathrm{d}^{3}y'\int\!\mathrm{d}^{3}z\int\!\mathrm{d}^{3}z'\,2N\left(\vec{x}'\right)\mathcal{G}_{ij,i'j'}\left(\vec{x}',\vec{z}\right)\frac{\delta^{2}S}{\delta K_{i'j'}\left(\vec{z}\right)\delta F\left(\vec{x}\right)}\nonumber \\
	&  & \qquad\times\frac{\delta^{2}S}{\delta h_{ij}\left(\vec{x}'\right)\delta K_{kl}\left(\vec{y}'\right)}\,\mathcal{G}_{kl,k'l'}\left(\vec{y}',\vec{z}'\right)\frac{\delta^{2}S}{\delta K_{k'l'}\left(\vec{z}'\right)\delta F\left(\vec{y}\right)}\nonumber \\
	&  & -\left(\vec{x}\leftrightarrow\vec{y}\right). \label{calF_def}
	\end{eqnarray}
When evaluating the functional derivatives, $N,h_{ij},F,K_{ij}$ are treated as independent.
Please note that by definition, $\mathcal{F}(\vec{x},\vec{y})$ is antisymmetric in the sense that $\mathcal{F}(\vec{x},\vec{y}) = - \mathcal{F}(\vec{y},\vec{x})$.

The derivation of the above two conditions is very involved, and interested readers may refer to Ref. \cite{Gao:2018znj} for details.
On the other hand, the meaning of the two conditions is clear according to the terminology of Dirac \cite{Henneaux:1992ig}: the degeneracy condition (\ref{degen_con}) implies a primary constraint, while the consistency condition guarantees that there is a secondary constraint associated with the primary one.
As long as the degeneracy condition (\ref{degen_con}) and the consistency condition (\ref{consis_con}) are satisfied, the action (\ref{S_scg}) describes theories that propagate at most 3 degrees of freedom.

\section{Perturbative analysis of the number of degrees of freedom} \label{sec:pert}

The purpose of this paper is to provide an alternative and also complementary understanding of the arising of the two conditions (\ref{degen_con}) and (\ref{consis_con}).

Let us consider the prototype action
	\begin{equation}
	S=\int\mathrm{d}t\mathrm{d}^{3}x\,N\sqrt{h}\left(a_{1}K+a_{2}F+b_{1}K_{ij}K^{ij}+b_{2}K^{2}+c_{1}KF+c_{2}F^{2}+\mathcal{V}\right),\label{S_quad_sim}
	\end{equation}
where coefficients $a_{i}$ and $b_{i}$ are general functions of $t,N,X$ with 
	\begin{equation}
		X\equiv \partial_{i}N\partial^{i}N.
	\end{equation}
In (\ref{S_quad_sim}), $\mathcal{V}$ stands for the ``potential'' terms that contain no time derivative, which do not affect the counting number of degrees of freedom.
In the following, we choose
	\begin{equation}
		\mathcal{V}=d_{1}+d_{2}R,
	\end{equation}
for concreteness.

The conditions for the action (\ref{S_quad_sim}) to have at most 3 degrees of freedom can be derived by evaluating the functional derivatives and plugging into (\ref{degen_con}) and (\ref{consis_con}), which have been shown in Ref. \cite{Gao:2018znj}. 
Here we simply summarize the final results and refer to Sec.\ref{sec:quad_cons} for a derivation of the more general case. The degeneracy condition for (\ref{S_quad_sim}) is 
	\begin{equation}
	 b_{1}\left[4\left(b_{1}+3b_{2}\right)c_{2}-3c_{1}^{2}\right]=0, \label{degen_sm}
	\end{equation}	
which implies $c_{2}$ is not independent and should be determined by
	\begin{equation}
	c_{2} = \frac{3}{4}\frac{c_{1}^{2}}{b_{1}+3b_{2}}.\label{c2_sol}
	\end{equation}
After plugging the solution for $c_{2}$ into (\ref{c2_sol}), the consistency condition implies two equations
	\begin{eqnarray}
	\frac{\partial a_{2}}{\partial X}-\frac{3c_{1}}{2(b_{1}+3b_{2})}\frac{\partial a_{1}}{\partial X} & = & 0, \label{cons_sm_1}\\
	\frac{\partial c_{1}}{\partial X}-\frac{c_{1}}{b_{1}+3b_{2}}\frac{\partial\left(b_{1}+3b_{2}\right)}{\partial X} & = & 0. \label{cons_sm_2}
	\end{eqnarray}
It is convenient to write
	\begin{eqnarray}
	c_{1} & \equiv & 2\beta\gamma, \label{gamma_beta_1}\\
	b_{1}+3b_{2} & \equiv & 3\beta, \label{gamma_beta_2}
	\end{eqnarray}
and thus (\ref{cons_sm_1}) and (\ref{cons_sm_2}) imply
	\begin{eqnarray}
	\gamma & = & \gamma\left(t,N\right)\neq0,\\
	a_{2}-\gamma\left(t,N\right)a_{1} & = & \alpha\left(t,N\right).
	\end{eqnarray}
There is no restriction to $\beta$. 	
In terms of $\beta$ and $\gamma$, the action with at most 3 degrees of freedom can be written as
	\begin{equation}
	S=\int\mathrm{d}t\mathrm{d}^{3}x\,N\sqrt{h}\left[a_{1}\left(K+\gamma F\right)+\alpha F+b_{1}\left(K_{ij}K^{ij}-\frac{1}{3}K^{2}\right)+\beta\left(K+\gamma F\right)^{2}+\mathcal{V}\right],\label{S_ori_fin}
	\end{equation}
where $a_{1},b_{1},\beta$ can be generally functions of $t,N,X$, while $\alpha,\gamma$ must be functions of $t$ and $N$ only. As for coefficients in the potential terms $\mathcal{V}$, there is no restriction.

\subsection{Linear perturbations around a homogeneous and isotropic background}

In the following, we examine how the degeneracy condition (\ref{degen_sm}) and the consistency condition (\ref{cons_sm_1}), (\ref{cons_sm_2}) for the action (\ref{S_quad_sim}) arise from the perturbative approach. 
This will provide us a complementary understanding of the two conditions.

We parametrize the metric to be
	\begin{equation}
	\mathrm{d}s^{2}\equiv-\left(e^{2A}-\mathfrak{g}^{ij}B_{i}B_{j}\right)\mathrm{d}t^{2}+2aB_{i}\,\mathrm{d}t\mathrm{d}x^{i}+a^{2}\mathfrak{g}_{ij}\mathrm{d}x^{i}\mathrm{d}x^{j},\label{metric_ADM}
	\end{equation}
which corresponds to the usual ADM variables as
\begin{eqnarray}
N & = & e^{A},\\
N_{i} & = & aB_{i},\\
h_{ij} & = & a^{2}\mathfrak{g}_{ij},
\end{eqnarray}
with $a=a\left(t\right)$ and $\mathfrak{g}^{ij}$ being the matrix inverse
of $\mathfrak{g}_{ij}$. 
A background is defined to be $A=0$, $B_{i}=0$, and $g_{ij} = \delta_{ij}$, which describes a homogeneous and isotropic universe.
It is convenient to define the perturbation of $\mathfrak{g}_{ij}$
in the exponential manner by denoting
\begin{equation}
\mathfrak{g}_{ij}\equiv\left(e^{\bm{H}}\right)_{ij}.\label{gij_eH}
\end{equation}
We split $H_{ij}$ into its trace part and traceless part as
\begin{equation}
H_{ij}=2\zeta\delta_{ij}+\hat{H}_{ij},\qquad\delta^{ij}\hat{H}_{ij}\equiv0,\label{Hij_split}
\end{equation}
and thus
\begin{equation}
\mathfrak{g}_{ij}=e^{2\zeta}\left(\delta_{ij}+\hat{H}_{ij}+\frac{1}{2}\hat{H}_{i}^{\phantom{i}k}\hat{H}_{kj}+\frac{1}{6}\hat{H}_{i}^{\phantom{i}k}\hat{H}_{kl}\hat{H}_{\phantom{l}j}^{l}+\mathcal{O}\left(\hat{H}^{4}\right)\right),\label{gij}
\end{equation}
where indices are raised/lowered by $\delta^{ij}$ and $\delta_{ij}$. As usual, we further decompose ($\partial^{2}\equiv\delta^{ij}\partial_{i}\partial_{j}$)
\begin{eqnarray}
B_{i} & \equiv & \partial_{i}B+S_{i},\label{Bi_dec}\\
\hat{H}_{ij} & \equiv & \left(\partial_{i}\partial_{j}-\frac{1}{3}\delta_{ij}\partial^{2}\right)E+\partial_{(i}F_{j)}+\gamma_{ij},\label{Hij_dec}
\end{eqnarray}
where $\zeta\equiv\frac{1}{6}\delta^{ij}H_{ij}$ is identified as
the scalar mode and $\gamma_{ij}$ is the tensor mode. With the above
settings
\begin{equation}
\sqrt{-g}\equiv N\sqrt{h}=a^{3}e^{A+3\zeta}.\label{detg_ADM}
\end{equation}

Straightforward expansion of the Lagrangian to the linear order in
perturbation variables $A$, etc., yields
\begin{equation}
S_{1}=\int\mathrm{d}t\mathrm{d}^{3}x \,a^{3}\left(\bar{\mathcal{E}}_{A}A+\bar{\mathcal{E}}_{\zeta}\,3\zeta\right),
\end{equation}
with
	\begin{eqnarray}
	\bar{\mathcal{E}}_{A} & = & -3H^{2}\left(b_{1}+3b_{2}-\frac{\partial\left(b_{1}+3b_{2}\right)}{\partial N}+3c_{1}\right)\nonumber \\
	&  & -3c_{1}\dot{H}-3H\left(-\frac{\partial a_{1}}{\partial N}+a_{2}\right)+d_{1}+\frac{\partial d_{1}}{\partial N}, \label{bgEoM_A}
	\end{eqnarray}
and
	\begin{equation}
	\bar{\mathcal{E}}_{\zeta}=d_{1}-3(b_{1}+3b_{2})H^{2}-2(b_{1}+3b_{2})\dot{H}. \label{bgEoM_z}
	\end{equation}
The background equations of motion are thus
\begin{equation}
\bar{\mathcal{E}}_{A}=0,\qquad\bar{\mathcal{E}}_{\zeta}=0.
\end{equation}

For the tensor modes, the quadratic-order action is 
	\begin{equation}
	S_{2}^{\mathrm{T}}=\int\mathrm{d}t\frac{\mathrm{d}^{3}k}{\left(2\pi\right)^{3}}\,\frac{1}{4}a^{3}\left(b_{1}\dot{\gamma}_{ij}\dot{\gamma}^{ij}-d_{2}\frac{k^{2}}{a^{2}}\gamma_{ij}\gamma^{ij}\right).
	\end{equation}
We must require
	\begin{equation}
	b_{1}>0,\qquad d_{2}>0,\label{ten_req}
	\end{equation}
in order to have no ghost and gradient instabilities.
Moreover, the propagation speed of the tensor modes is given by
	\begin{equation}
		c_{\mathrm{T}}^{2}=\frac{d_{2}}{b_{1}}. 
	\end{equation}
As has been discussed in Refs. \cite{Lombriser:2015sxa,Lombriser:2016yzn,McManus:2016kxu}, the propagation speed of the gravitational waves may put stringent constraint on scalar-tensor theories. The detection of GW170817 \cite{TheLIGOScientific:2017qsa} and GRB170817A \cite{Monitor:2017mdv} indicates that the propagation speed of the gravitational waves coincides with the speed of
light with deviations smaller than approximately $10^{-15}$. 
Although the physics of GW170817 may be different from that in the primordial universe, it has already been used to restrict the structure of scalar-tensor theories \cite{Creminelli:2017sry,Sakstein:2017xjx,Ezquiaga:2017ekz,Baker:2017hug,Langlois:2017dyl}. 
For the simple model we are considering, this implies that $b_{1} = d_{2}$ at the background level.

The quadratic action for the scalar modes takes the general form
	\begin{equation}
	S_{2}^{\mathrm{S}}\left[\zeta,A,B\right]=\int\mathrm{d}t\mathrm{d}^{3}x\,a^{3}\left(L_{2}^{(1)}+L_{2}^{(2)}\right),\label{S2_S_ori}
	\end{equation}
where 
	\begin{eqnarray}
	L_{2}^{(1)} & = & \mathcal{C}_{\dot{\zeta}^{2}}\dot{\zeta}^{2}+\mathcal{C}_{\dot{\zeta}\dot{A}}\dot{\zeta}\dot{A}+\mathcal{C}_{\dot{A}^{2}}\dot{A}^{2}\nonumber \\
	&  & -\mathcal{C}_{\dot{\zeta}B}\dot{\zeta}\frac{\partial^{2}B}{a} -\mathcal{C}_{\dot{A}B}\dot{A}\frac{\partial^{2}B}{a}+\mathcal{C}_{B^{2}}\frac{(\partial^{2}B)^{2}}{a^{2}},
	\end{eqnarray}
which are terms relevant to counting the number of degrees of freedom, with 
	\begin{eqnarray}
	\mathcal{C}_{\dot{\zeta}^{2}} & = & 3(b_{1}+3b_{2}),\\
	\mathcal{C}_{\dot{\zeta}\dot{A}} &= & 3c_{1},\\
	\mathcal{C}_{\dot{A}^{2}} & = & c_{2},\\
	\mathcal{C}_{\dot{\zeta}B} & = & 2(b_{1}+3b_{2}),\\
	\mathcal{C}_{\dot{A}B} & = & c_{1},\\
	\mathcal{C}_{B^{2}} & = & b_{1}+b_{2}.
	\end{eqnarray}
Terms which are irrelevant to counting the number of degrees of freedom are
	\begin{eqnarray}
	L_{2}^{(2)} & = & \mathcal{C}_{\zeta^{2}}\zeta^{2}+\mathcal{C}_{\dot{\zeta}A}\dot{\zeta}A+\mathcal{C}_{\zeta A}\zeta A+\mathcal{C}_{A^{2}}A^{2} - \mathcal{C}_{AB}A\frac{\partial^{2} B}{a}, \label{L2^2}
	\end{eqnarray}
in which various coefficients are given in Appendix \ref{app:coeff}.

Throughout this paper, we assume $ b_1+3b_2 \neq 0$.

\subsection{Degeneracy condition} \label{sec:degen}

In the quadratic action (\ref{S2_S_ori}), $B$ behaves as an auxiliary variable (i.e., acquires no time derivative). 
According to whether $\mathcal{C}_{B^{2}}\equiv b_{1}+b_{2}$ is vanishing or not, we have to discuss two cases.

First, if $\mathcal{C}_{B^{2}}\equiv b_{1}+b_{2}=0$ (as in the case of GR), $B$ enters the quadratic Lagrangian linearly and serves as a Lagrange multiplier.
In this case, from (\ref{S2_S_ori}), the equation of motion for $B$ is
	\begin{equation}
	\mathcal{C}_{\dot{\zeta}B}\dot{\zeta}+\mathcal{C}_{\dot{A}B}\dot{A}+\mathcal{C}_{AB}A=0.\label{Beq_lm}
	\end{equation}
In the case $b_{1}+3b_{2}\neq 0$, Eq. (\ref{Beq_lm}) is a nonholonomic constraint between $\zeta$ and $A$, which by itself does not kill any degree of freedom. 
As a result, in this case, reducing the number of degrees of freedom requires the degeneracy of the the kinetic terms of the residual variables (i.e., $\zeta$ and $A$)
	\begin{equation}
	\det\left(\begin{array}{cc}
	\mathcal{C}_{\dot{\zeta}^{2}} & \frac{1}{2}\mathcal{C}_{\dot{\zeta}\dot{A}}\\
	\frac{1}{2}\mathcal{C}_{\dot{\zeta}\dot{A}} & \mathcal{C}_{\dot{A}^{2}}
	\end{array}\right)=3(b_{1}+3b_{2})c_{2}-\frac{9}{4}c_{1}^{2}=0. \label{degen_gr}
	\end{equation}
This is nothing but (\ref{degen_sm}).
	
If $\mathcal{C}_{B^{2}}\equiv b_{1}+b_{2} \neq 0$, the equation of motion for $B$ is
	\begin{equation}
	\mathcal{C}_{\dot{\zeta}B}\dot{\zeta}+\mathcal{C}_{\dot{A}B}\dot{A}+\mathcal{C}_{AB}A-2\mathcal{C}_{B^{2}}\frac{\partial^{2}B}{a}=0,\label{Beq}
	\end{equation}
from which we may solve $B$ formally:
	\begin{equation}
	\partial^{2}\frac{B}{a} = \frac{1}{2\mathcal{C}_{B^{2}}}\left(\mathcal{C}_{\dot{\zeta}B}\dot{\zeta}+\mathcal{C}_{\dot{A}B}\dot{A}+\mathcal{C}_{AB}A\right).\label{B_sol_form}
	\end{equation}
Plugging the solution (\ref{B_sol_form}) into (\ref{S2_S_ori}) yields the action for $\zeta$ and $A$,
	\begin{eqnarray}
	S_{2}^{\mathrm{S}}\left[\zeta,A\right] & \equiv & \int\mathrm{d}t\frac{\mathrm{d}^{3}k}{\left(2\pi\right)^{3}}a^{3}\,\Big[\mathcal{D}_{\dot{\zeta}^{2}}\dot{\zeta}^{2}+\mathcal{D}_{\dot{\zeta}\dot{A}}\dot{\zeta}\dot{A}+\mathcal{D}_{\dot{A}^{2}}\dot{A}^{2}\nonumber \\
	&  & +\mathcal{C}_{\zeta^{2}}\zeta^{2}+\mathcal{D}_{\dot{\zeta}A}\dot{\zeta}A+\mathcal{C}_{\zeta A}\zeta A+\mathcal{D}_{A^{2}}A^{2}\Big],\label{S2_zeta_A}
	\end{eqnarray}
in which the new coefficients are given by
	\begin{eqnarray}
	\mathcal{D}_{\dot{\zeta}^{2}} & = & \mathcal{C}_{\dot{\zeta}^{2}}-\frac{1}{4}\frac{\mathcal{C}_{\dot{\zeta}B}^{2}}{\mathcal{C}_{B^{2}}},\label{calD_zd2}\\
	\mathcal{D}_{\dot{\zeta}\dot{A}} & = & \mathcal{C}_{\dot{\zeta}\dot{A}}-\frac{1}{2}\frac{\mathcal{C}_{\dot{\zeta}B}\mathcal{C}_{\dot{A}B}}{\mathcal{C}_{B^{2}}},\\
	\mathcal{D}_{\dot{A}^{2}} & = & \mathcal{C}_{\dot{A}^{2}}-\frac{1}{4}\frac{\mathcal{C}_{\dot{A}B}^{2}}{\mathcal{C}_{B^{2}}},\label{calD_Ad2}
	\end{eqnarray}
	and
	\begin{eqnarray}
	\mathcal{D}_{\dot{\zeta}A} & = & \mathcal{C}_{\dot{\zeta}A}-\frac{1}{2}\frac{\mathcal{C}_{\dot{\zeta}B}\mathcal{C}_{AB}}{\mathcal{C}_{B^{2}}},\\
	\mathcal{D}_{A^{2}} & = & \mathcal{C}_{A^{2}}-\frac{1}{4}\frac{\mathcal{C}_{AB}^{2}}{\mathcal{C}_{B^{2}}}+\frac{1}{4}\frac{1}{a^{3}}\partial_{t}\left(a^{3}\frac{\mathcal{C}_{AB}\mathcal{C}_{\dot{A}B}}{\mathcal{C}_{B^{2}}}\right).
	\end{eqnarray}	
The coefficients for the kinetic terms can be evaluated explicitly to be
	\begin{eqnarray}
	\mathcal{D}_{\dot{\zeta}^{2}} & = & \frac{2b_{1}(b_{1}+3b_{2})}{b_{1}+b_{2}},\\
	\mathcal{D}_{\dot{\zeta}\dot{A}} & = & \frac{2b_{1}c_{1}}{b_{1}+b_{2}},\\
	\mathcal{D}_{\dot{A}^{2}} & = & c_{2}-\frac{c_{1}^{2}}{4(b_{1}+b_{2})}.
	\end{eqnarray}
According to the above results, we have seen clearly that generally both $\zeta$ and $A$ acquire kinetic terms and are dynamical, which implies that there are two scalar modes propagating in the theory (at quadratic order around an FRW background).  
	
To be more precise, let us check the Hessian matrix of the kinetic terms, which is given by
	\begin{equation}
	\det\left(\begin{array}{cc}
	\mathcal{D}_{\dot{\zeta}^{2}} & \frac{1}{2}\mathcal{D}_{\dot{\zeta}\dot{A}}\\
	\frac{1}{2}\mathcal{D}_{\dot{\zeta}\dot{A}} & \mathcal{D}_{\dot{A}^{2}}
	\end{array}\right)=\frac{b_{1}}{2(b_{1}+b_{2})}\left[4(b_{1}+3b_{2})c_{2}-3c_{1}^{2}\right].\label{detkin}
	\end{equation}
If the Hessian matrix does not degenerate, both $\zeta$ and $A$ are dynamical.
At this point, we emphasize that the theory that possesses two scalar modes does not necessarily mean it is pathological.
In the case of generally covariant scalar-tensor theories with higher derivatives, the extra scalar mode is associated with the Ostrogradsky instability due to the higher time derivatives.
In the framework of the general action (\ref{S_scg}), however, since the action involves only up to the first-order time derivatives, there is no Ostrogradsky ghost at all. We may claim that the theory is healthy if the two scalar modes are both well behaved, e.g., with correct signs of kinetic terms and without gradient instability.
Nevertheless, in the present work, we focus on the case with a single scalar degree of freedom for the following reasons.
First, when apparently recovering the general covariance using the Stueckelberg trick, having a single scalar mode in the form of spatially covariant gravity (i.e., unitary gauge) is a necessary condition to evade the Ostrogradsky ghost in the generally covariant formalism.
Second, many models of inflation and dark energy involve a single scalar field. An adiabatic initial condition from single field/clock models of inflation is much preferred by the current observations.
For our purpose to have a single scalar degree of freedom, we need to require the kinetic terms to be degenerate. From (\ref{detkin}) and keeping in mind that $b_{1}>0$ in order to have healthy tensor perturbations, it implies
	\begin{equation}
	4(b_{1}+3b_{2})c_{2}-3c_{1}^{2}=0.\label{degen}
	\end{equation}
Equation (\ref{degen}) is nothing but the same condition as (\ref{degen_gr}).

After imposing the degeneracy condition, the kinetic terms in (\ref{S2_zeta_A}) become a perfect square trinomial
	\begin{eqnarray}
	&  & \mathcal{D}_{\dot{\zeta}^{2}}\dot{\zeta}^{2}+\mathcal{D}_{\dot{\zeta}\dot{A}}\dot{\zeta}\dot{A}+\mathcal{D}_{\dot{A}^{2}}\dot{A}^{2}\nonumber \\
	& = & \mathcal{D}_{\dot{\zeta}^{2}}\left[\dot{\zeta}+\frac{c_{1}}{2(b_{1}+3b_{2})}\dot{A}\right]^{2}.\label{kin_ps}
	\end{eqnarray}
Thus we may introduce a new variable,
	\begin{equation}
	\tilde{\zeta} :=\zeta+\frac{1}{3}\gamma\,A, \label{zeta_tld_1st}
	\end{equation}
where we denote
	\begin{equation}
	\gamma=\frac{3c_{1}}{2\left(b_{1}+3b_{2}\right)}, \label{gamma_def}
	\end{equation}
for short and for later convenience.
In terms of the new variable $\tilde{\zeta}$, the quadratic action is
	\begin{eqnarray}
	S_{2}^{\mathrm{S}}\left[\tilde{\zeta},A\right] & = & \int\mathrm{d}t\frac{\mathrm{d}^{3}k}{\left(2\pi\right)^{3}}a^{3}\,\Big(\mathcal{D}_{\dot{\zeta}^{2}}\dot{\tilde{\zeta}}^{2}+\mathcal{C}_{\zeta^{2}}\tilde{\zeta}^{2}\nonumber \\
	&  & +\mathcal{F}_{\dot{\zeta}A}\dot{\tilde{\zeta}}A+\mathcal{F}_{\zeta A}\tilde{\zeta}A+\mathcal{F}_{A^{2}}A^{2}\Big),\label{S2_zetah_A}
	\end{eqnarray}
in which it is transparent that only $\tilde{\zeta}$ acquires dynamics, while $A$ becomes an auxiliary variable.
Coefficients in (\ref{S2_zetah_A}) are given in Appendix \ref{app:effact}.
It is thus a standard exercise to solve the auxiliary variable $A$ and derive the final action for the single variable $\tilde{\zeta}$, which can be also found in Appendix \ref{app:effact}.

To end this section, we conclude that, as long as the degeneracy condition (\ref{degen}) is satisfied, our model propagates a single scalar mode at the linear order, when expanded around a FRW background.

\section{Consistency condition} \label{sec:cons}

According to the analysis in the previous section, in order to have a single scalar mode at the linear order when expanding around a homogeneous and isotropic background, the degeneracy condition (\ref{degen}) is the only condition that must be imposed. One may be curious as to how the consistency condition (\ref{cons_sm_1}), (\ref{cons_sm_2}) arises.
As we shall show in this section, if only the degeneracy condition is satisfied, generally the unwanted scalar mode will reappear, either at nonlinear orders when expanding around an FRW background or at the linear order when expanding around an inhomogeneous background.
In both cases, the same consistency condition must be imposed in addition to the degeneracy condition in order to evade the unwanted mode.

\subsection{Cubic-order perturbations around the FRW background}

First, let us examine what happens when going to higher orders around a FRW background. 
Generally the scalar modes and tensor modes get coupled on nonlinear order. For our purpose, we focus on the pure scalar sector.
Straightforward expansion of the action (\ref{S_quad_sim}) yields the cubic action for the scalar modes
	\begin{equation}
	S_{3}^{\mathrm{S}}\left[\zeta,A,B\right] = \int \mathrm{d}t\mathrm{d}^3 x\,\mathcal{L}_{3}^{\mathrm{S}}\left(\zeta,A,B\right), \label{S3_zetaAB}
	\end{equation}
where the explicit expression for the cubic Lagrangian $\mathcal{L}_{3}^{\mathrm{S}}$ can be found in Appendix \ref{app:L3} due to its length.
For our purpose, we simply need to focus on terms which are relevant to the number of degrees of freedom.
According to (\ref{L3_zetaAB}), the existence of the terms
	\begin{eqnarray}
	\mathcal{L}_{3}^{\mathrm{S}}\left(\zeta,A,B\right) & \supset & \partial_{i}A\partial^{i}A\dot{A}\,\frac{a}{2}\frac{\partial\left(a_{2}+3c_{1}H\right)}{\partial X}-A\dot{A}\partial^{2}B\,a^{2}\frac{\partial c_{1}}{\partial N}\nonumber \\
	&  & -\zeta\dot{A}\partial^{2}B\,a^{2}c_{1}-\partial_{i}B\partial^{i}A\dot{A}\,2a^{2}c_{2}-\partial_{i}\zeta\partial^{i}B\dot{A}\,a^{2}c_{1}\nonumber \\
	&  & +A\dot{A}^{2}a^{3}\left(c_{2}+\frac{\partial c_{2}}{\partial N}\right)+\zeta\dot{A}^{2}3a^{3}c_{2}\nonumber \\
	&  & +A\dot{A}\dot{\zeta}\,3a^{3}\frac{\partial c_{1}}{\partial N}+\dot{A}\zeta\dot{\zeta}\,9a^{3}c_{1}, \label{L3_dger}
	\end{eqnarray}
explicitly prevent $A$ from being an auxiliary variable at the cubic order, since the time derivative of $A$ cannot be removed by integrations by parts, without introducing second-order time derivatives of $\zeta$.
It is also transparent that if the original action contains no $F  \propto \dot{N}$ terms (such as SCG in Refs. \cite{Gao:2014soa,Gao:2014fra}) $a_{2}=c_{1} = c_{2} \equiv 0$, and all terms in (\ref{L3_dger}) vanish. As a result, $A$ keeps serving as an auxiliary variable up to the cubic order in theories in Refs. \cite{Gao:2014soa,Gao:2014fra}.
In fact, as has been proven in Refs. \cite{Gao:2014fra} and recently in a more general setting \cite{Gao:2018znj}, as long as the action contains no $F$ terms, the theory propagates at most 3 degrees of freedom up to arbitrarily high order or, precisely, in a nonperturbative sense.

According to the previous analysis, as long as the degeneracy condition (\ref{degen}) is satisfied, by employing the new variables $\{\tilde{\zeta},A\}$, $A$ becomes an auxiliary variable explicitly, and there is only a single scalar mode $\tilde{\zeta}$ propagating at the quadratic order.
Thus one may expect that at the cubic-order the situation may get cured by the same operation.
By replacing $\zeta$ in terms of $\tilde{\zeta}$ defined in (\ref{zeta_tld_1st}) and after a tedious manipulation, we get the cubic order action for $\{\tilde{\zeta},A,B\}$. Because of its length, we tend not to present the full expressions in the present work.
Instead, we pay special attention to terms that are potentially dangerous:
	\begin{eqnarray}
	\mathcal{L}_{3}\left(\tilde{\zeta},A,B\right) & \supset & \dot{A}\partial_{i}A\partial^{i}A\,\frac{a}{4}\left[2\frac{\partial a_{2}}{\partial X}-\frac{3c_{1}}{b_{1}+3b_{2}}\frac{\partial a_{1}}{\partial X}+6\left(\frac{\partial c_{1}}{\partial X}-\frac{c_{1}}{b_{1}+3b_{2}}\frac{\partial(b_{1}+3b_{2})}{\partial X}\right)H\right]\nonumber \\
	&  & +A\dot{A}\left(\dot{\tilde{\zeta}}\,3a^{3}-\partial^{2}B\,a^{2}\right)\left[c_{1}+\frac{\partial c_{1}}{\partial N}-\frac{c_{1}}{b_{1}+3b_{2}}\frac{\partial(b_{1}+3b_{2})}{\partial N}\right]. \label{L3_naive_dg}
	\end{eqnarray}
The existence of these terms implies that simply replacing $\zeta$ in terms of $\tilde{\zeta}$ defined in (\ref{zeta_tld_1st}) is not sufficient to remove the time derivative terms of $A$ at the cubic order.

At this point, one may conclude that we have to impose further conditions [besides the degeneracy condition (\ref{degen})] by requiring that terms in both square brackets in (\ref{L3_naive_dg}) must be vanishing identically.
This, however, turns out to be too strong, since the second line in (\ref{L3_naive_dg}) can be removed by observing that it has the same structure of the terms approximately $\dot{\tilde{\zeta}} \left(\dot{\tilde{\zeta}}\,3a^{3}-\partial^{2}B\,a^{2}\right)$ in the quadratic action $\mathcal{L}_{2}\left(\tilde{\zeta},A,B\right)$ with new variable $\tilde{\zeta}$.
This observation indicates that we have to extend the definition of the new variable $\tilde{\zeta}$ to the second order, which is proven to be
	\begin{equation}
	\tilde{\zeta}=\zeta+\gamma\,A+\frac{1}{2}\left(\gamma+\frac{\partial\gamma}{\partial N}\right)A^{2}, \label{zeta_tld_2nd}
	\end{equation}
with $\gamma$ being the same as in (\ref{gamma_def}).
Please note that by employing (\ref{zeta_tld_2nd}) the quadratic-order action $S_{2}^{\mathrm{S}}\left[\zeta,A,B\right]$ also contributes to the cubic-order action $S_{3}^{\mathrm{S}}\left[\tilde{\zeta},A,B\right]$.
In fact, these additional contributions from $S_{2}^{\mathrm{S}}\left[\zeta,A,B\right]$ exactly cancel the second term in (\ref{L3_naive_dg}), leaving us only the first term in (\ref{L3_naive_dg}).

After using (\ref{zeta_tld_2nd}), the second line in (\ref{L3_naive_dg}) gets canceled, and we are left with only one pathological term proportional to
	\begin{equation}
		\dot{A}\partial_{i}A\partial^{i}A.
	\end{equation}
Such a term yields an equation of motion for $A$ that is first order in the time derivative, which signals the existence of half a degree of freedom \cite{Henneaux:2009zb,Li:2009bg,Blas:2009yd}.
In order to prevent such a situation, the coefficient must be vanishing identically, that is
	\begin{equation}
		2\frac{\partial a_{2}}{\partial X}-\frac{3c_{1}}{b_{1}+3b_{2}}\frac{\partial a_{1}}{\partial X}+6\left(\frac{\partial c_{1}}{\partial X}-\frac{c_{1}}{b_{1}+3b_{2}}\frac{\partial(b_{1}+3b_{2})}{\partial X}\right)H= 0. \label{L3_con}
	\end{equation}
To have (\ref{L3_con}) be valid all the time, we must require
	\begin{eqnarray}
	2\frac{\partial a_{2}}{\partial X}-\frac{3c_{1}}{b_{1}+3b_{2}}\frac{\partial a_{1}}{\partial X} & = & 0, \label{cons_sm_1a}\\
	\frac{\partial c_{1}}{\partial X}-\frac{c_{1}}{b_{1}+3b_{2}}\frac{\partial(b_{1}+3b_{2})}{\partial X} & = & 0, \label{cons_sm_2a}
	\end{eqnarray}
to be satisfied separately, which are nothing but the equations required by the consistency condition (\ref{cons_sm_1})-(\ref{cons_sm_2}).

It is also interesting that the requirement of the consistency condition appears as early as at the cubic order. In other words, for the prototype action (\ref{S_quad_sim}), if one is able to evade the unwanted scalar mode up to the cubic order when expanding around a FRW background, the theory contains no unwanted mode up to arbitrarily high orders in perturbations.

\subsection{Quadratic-order perturbations around an inhomogeneous background}

For an alternative but also complementary analysis, let us consider the linear perturbations around an inhomogeneous background.
We make the ansatz for the inhomogeneous background metric, following the same notations as in ref. \cite{Gao:2018qpy},
\begin{equation}
\mathrm{d}s^{2}=-\bar{N}^{2}\mathrm{d}t^{2}+a^{2}\bar{\mathfrak{g}}_{ij}\mathrm{d}x^{i}\mathrm{d}x^{j},
\end{equation}
where $a=a\left(t\right)$ is the standard scale factor, and 
	\begin{eqnarray}
	\bar{N} & = & \bar{N}\left(\vec{x}\right),\\
	\bar{\mathfrak{g}}_{ij} & = & \bar{\mathfrak{g}}_{ij}\left(\vec{x}\right),
	\end{eqnarray}
which are functions of space coordinates only.
The metric is parametrized the same as (\ref{metric_ADM}), that is,
\begin{eqnarray}
N & = & \bar{N}e^{A},\\
N_{i} & = & aB_{i},\\
h_{ij} & = & a^{2}\mathfrak{g}_{ij}.
\end{eqnarray}
We define the perturbation of the spatial metric $\mathfrak{g}_{ij}$
through
\begin{equation}
\mathfrak{g}_{ij}\equiv\bar{\mathfrak{g}}_{ik}\left(e^{\bm{H}}\right)_{\phantom{k}j}^{k},\label{gij_eH_inhomo}
\end{equation}
and the inverse is simply
\begin{equation}
\mathfrak{g}^{ij}=\left(e^{-\bm{H}}\right)_{\phantom{i}k}^{i}\bar{\mathfrak{g}}^{kj}.
\end{equation}
Similar to (\ref{Hij_split}), we split $H_{ij}$ into
\begin{equation}
H_{ij}=2\zeta\,\bar{\mathfrak{g}}_{ij}+\hat{H}_{ij},
\end{equation}
where $\hat{H}_{ij}$ is traceless satisfying $\bar{\mathfrak{g}}^{ij}\hat{H}_{ij}=0$.
As usual, we further decompose ($\bar{\nabla}^{2}\equiv\bar{g}^{ij}\partial_{i}\partial_{j}$)
\begin{eqnarray}
B_{i} & \equiv & \partial_{i}B+S_{i},\label{Bi_dec-1}\\
\hat{H}_{ij} & \equiv & \left(\bar{\nabla}_{i}\bar{\nabla}_{j}-\frac{1}{3}\bar{\mathfrak{g}}_{ij}\bar{\nabla}^{2}\right)E+\bar{\nabla}_{(i}F_{j)}+\gamma_{ij}.\label{Hij_dec_inhomo}
\end{eqnarray}
For our purpose, we focus on the scalar modes only and choose the gauge $E=0$ and $F_{i}=0$; thus, $\hat{H}_{ij}\equiv0$.

After tedious but straightforward manipulations, we have derived the quadratic-order action for perturbations $A$, $B$, and $\zeta$.
Again, we tend not to present the full expression in the current paper, due to its length.
Instead, we focus on terms which prevent $A$ from being an auxiliary variable.
The relevant terms in the quadratic Lagrangian are
	\begin{eqnarray}
	\mathcal{L}_{2}(A,B,\zeta) & \supset & -c_{1}a^{2}\bar{\nabla}^{2}B\dot{A}-2c_{2}a^{2}\partial_{i}\bar{N}\partial^{i}B\dot{A}\nonumber \\
	&  & +\bar{N}a\left(\frac{\partial a_{2}}{\partial X}\bar{N}+3H\frac{\partial c_{1}}{\partial X}\right)\partial_{i}\bar{N}\partial^{i}A\,\dot{A}\nonumber \\
	&  & +2c_{2}a^{3}\bar{N}\dot{A}^{2}+6c_{1}a^{3}\dot{A}\dot{\zeta}+6(b_{1}+3b_{2})\frac{a^{3}}{\bar{N}}\dot{\zeta}^{2},
	\end{eqnarray}
where we keep the last term since it will contribute $\dot{A}$ after the replacement (\ref{zeta_tld_1st}).
An analysis parallel to that in Sec.\ref{sec:degen} implies the same degeneracy condition (\ref{degen}). After making the same replacement (\ref{zeta_tld_1st}), we are left with only one term that is dangerous:
	\begin{eqnarray}
	\mathcal{L}_{2}(A,B,\tilde{\zeta}) & \supset & \frac{1}{2}\dot{A}\partial_{i}A\partial^{i}\bar{N}\,\bar{N}a\bigg\{\left(2\frac{\partial a_{2}}{\partial X}-\frac{3c_{1}}{b_{1}+3b_{2}}\frac{\partial a_{1}}{\partial X}\right)\bar{N}\nonumber \\
	&  & \qquad+6H\left(\frac{\partial c_{1}}{\partial X}-\frac{c_{1}}{b_{1}+3b_{2}}\frac{\partial(b_{1}+3b_{2})}{\partial X}\right)\bigg\}. \label{dgr_inh}
	\end{eqnarray}
The above term is proportional to
	\begin{equation}
		\dot{A}\partial_{i}A\partial^{i}\bar{N},
	\end{equation}
in which the time derivative of $A$ cannot be removed by integrations by parts. Again, the presence of such a term prevents $A$ from being an auxiliary variable.
To get rid of such a term, one needs to require terms in the square bracket in (\ref{dgr_inh}) to be vanishing identically. This reproduces exactly the same consistency condition (\ref{cons_sm_1a}), (\ref{cons_sm_2a}).

The above result should not be strange, since nonlinear perturbations around a homogeneous and isotropic background are highly related to linear perturbations around an inhomogeneous background.

\section{Quadratic construction and field transformations} \label{sec:quad_trans}

As has been shown in Ref. \cite{Gao:2018znj}, for action taking the general form (\ref{S_scg}), as long as the degeneracy and consistency conditions (\ref{degen_con}) and (\ref{consis_con}) are satisfied, there are at most 3 physical degrees of freedom.
Comparing with spatially covariant theories without $F$ terms which are always ``healthy,'' one would be interested in the relation between theories with  and those without $F$ terms.
In particular, the introduction of the new variable $\tilde{\zeta}$ indicates that healthy theories with $F$ terms may be related to theories without $F$ terms through some redefinition of variables.
In this section, we study this issue by constructing a healthy theory with $F$ terms explicitly.

\subsection{Quadratic construction} \label{sec:quad_cons}

Let us consider the action
	\begin{eqnarray}
	S^{\mathrm{(quad)}} & = & \int\mathrm{d}t\mathrm{d}^{3}x\,N\sqrt{h}\Big(a_{1}K+a_{2}F+a_{3}X^{ij}K_{ij}\nonumber \\
	&  & \qquad\qquad+b_{1}K_{ij}K^{ij}+b_{2}K^{2}+b_{3}X^{ij}K_{ij}K+b_{4}X^{ij}K_{ik}K_{\phantom{k}j}^{k}+b_{5}\left(X^{ij}K_{ij}\right)^{2}\nonumber \\
	&  & \qquad\qquad+c_{1}KF+c_{2}F^{2}+c_{3}X^{ij}K_{ij}F+\mathcal{V}\Big), \label{S_quad}
	\end{eqnarray}
where 
	\begin{equation}
	X_{ij}\equiv\frac{1}{2}\partial_{i}N\partial_{j}N, \label{X_ij_def}
	\end{equation}
and various coefficients $a_{1}$, etc., are general functions of $t,N,X$ with $X=h^{ij}X_{ij}$.
The action (\ref{S_quad}) is written such that it is the most general action quadratic in $K_{ij}$ and $F$ and also contains up to the quadratic order in $X_{ij}$.
Again, in (\ref{S_quad}), $\mathcal{V}$ stands for terms without time derivatives, which are irrelevant to counting number of degrees of freedom so that we do not give the explicit assumption of their expressions.

The second-order functional derivatives are given by 
	\begin{eqnarray}
	\frac{\delta^{2}S}{\delta F\left(\vec{x}\right)\delta F\left(\vec{y}\right)} & = & \delta^{3}\left(\vec{x}-\vec{y}\right)N\sqrt{h}\,2c_{2},\label{d2S_AA}\\
	\frac{\delta^{2}S}{\delta K_{ij}\left(\vec{x}\right)\delta F\left(\vec{y}\right)} & = & \delta^{3}\left(\vec{x}-\vec{y}\right)N\sqrt{h}\left(c_{1}h^{ij}+c_{3}X^{ij}\right),\label{d2S_AB}\\
	\frac{\delta^{2}S}{\delta K_{ij}\left(\vec{x}\right)\delta K_{kl}\left(\vec{y}\right)} & = & \delta^{3}\left(\vec{x}-\vec{y}\right)\Delta^{ij,kl}\left(\vec{x}\right),\label{d2S_BB}
	\end{eqnarray}
with
	\begin{eqnarray}
	\Delta^{ij,kl} & := & N\sqrt{h}\bigg[2b_{1}\mathbf{I}^{ij,kl}+2b_{2}h^{ij}h^{kl}+b_{3}\left(h^{ij}X^{kl}+h^{kl}X^{ij}\right)\nonumber \\
	&  & \qquad+b_{4}\frac{1}{2}\left(h^{ik}X^{jl}+h^{il}X^{jk}+h^{jk}X^{il}+h^{jl}X^{ik}\right)+2b_{5}X^{ij}X^{kl}\bigg].\label{Dlt^ijkl}
	\end{eqnarray}
We make the ansatz for the inverse of $\delta^{3}\left(\vec{x}-\vec{y}\right)\Delta^{ij,kl}\left(\vec{x}\right)$,
	\begin{equation}
	\mathcal{G}_{ij,kl}\left(\vec{x},\vec{y}\right)=\delta^{3}\left(\vec{x}-\vec{y}\right)\mathcal{G}_{ij,kl}\left(\vec{x}\right),
	\end{equation}
with
	\begin{eqnarray}
	\mathcal{G}_{ij,kl} & = & \frac{1}{N\sqrt{h}}\bigg[\frac{1}{2b_{1}}\mathbf{I}_{ij,kl}+\alpha\,h_{ij}h_{kl}+\beta\left(h_{ij}X_{kl}+h_{kl}X_{ij}\right)\nonumber \\
	&  & \qquad+\gamma\left(h_{ik}X_{jl}+h_{il}X_{jk}+h_{jk}X_{il}+h_{jl}X_{ik}\right)+\rho\,X^{ij}X^{kl}\bigg], \label{calG_ans}
	\end{eqnarray}
where $\alpha,\beta,\gamma,\rho$ are coefficients to be determined.
In fact, due to the ``bivector'' form of $X_{ij}$ [see (\ref{X_ij_def})], Eq. (\ref{calG_ans}) is the most general expression one can write.
The coefficients $\alpha,\beta,\gamma,\rho$ can be determined from the definition (\ref{calG_inv}), which is now simplified to be
	\begin{equation}
	\Delta^{ij,mn}\mathcal{G}_{mn,kl}=\mathbf{I}_{kl}^{ij}. \label{inv_eq}
	\end{equation}
There is a unique set of solutions for (\ref{inv_eq}),
	\begin{eqnarray}
	\alpha & = & \frac{1}{4b_{1}\Xi}\left[-4b_{1}b_{2}-4b_{2}b_{4}X+\left(b_{3}^{2}-4b_{2}b_{5}\right)X^{2}\right],\\
	\beta & = & \frac{1}{4b_{1}\Xi}\left[-2b_{1}b_{3}+4b_{2}b_{4}+\left(4b_{2}b_{5}-b_{3}^{2}\right)X\right],\\
	\gamma & = & -\frac{b_{4}}{4b_{1}\left(2b_{1}+b_{4}X\right)},\\
	\rho & = & \frac{1}{4b_{1}\left(2b_{1}+b_{4}X\right)\Xi}\big[-8b_{5}b_{1}^{2}+\left(6b_{3}^{2}+8b_{4}b_{3}+4\left(b_{4}^{2}-6b_{2}b_{5}\right)\right)b_{1}\nonumber \\
	&  & \qquad\qquad+4b_{2}b_{4}^{2}+b_{4}\left(4\left(b_{1}+b_{2}\right)b_{5}-b_{3}^{2}\right)X\big],
	\end{eqnarray}
where we define
	\begin{equation}
	\Xi:=\left[2\left(b_{1}+2b_{2}\right)b_{5}-b_{3}^{2}\right]X^{2}+\left[4b_{2}b_{4}+2b_{1}\left(b_{3}+b_{4}\right)\right]X+2b_{1}\left(b_{1}+3b_{2}\right),\label{Xi_def}
	\end{equation}
for shorthand. We must require $\Xi\neq0$, which is to require that the kinetic terms for $K_{ij}$ do not degenerate.

Now we are ready to write the degeneracy and consistency conditions.
After some manipulations, $\mathcal{D}\left(\vec{x},\vec{y}\right)$ defined in (\ref{degen_con}) takes the form
	\begin{equation}
	\mathcal{D}\left(\vec{x},\vec{y}\right)\equiv\delta^{3}\left(\vec{x}-\vec{y}\right)N\sqrt{h}\,\frac{1}{\Xi}\left(\mathcal{D}_{0}+\mathcal{D}_{1}X+\mathcal{D}_{2}X^{2}\right), \label{con_degen}
	\end{equation}	
with
	\begin{equation}
	\mathcal{D}_{0}=b_{1}\left[4\left(b_{1}+3b_{2}\right)c_{2}-3c_{1}^{2}\right],
	\end{equation}
	\begin{equation}
	\mathcal{D}_{1}=4b_{1}b_{3}c_{2}+2b_{4}\left(2\left(b_{1}+2b_{2}\right)c_{2}-c_{1}^{2}\right)-2b_{1}c_{1}c_{3},
	\end{equation}
	\begin{eqnarray}
	\mathcal{D}_{2} & = & -2b_{3}^{2}c_{2}+2b_{3}c_{1}c_{3}-\left(b_{1}+2b_{2}\right)c_{3}^{2}\nonumber \\
	&  & +2b_{5}\left(2\left(b_{1}+2b_{2}\right)c_{2}-c_{1}^{2}\right),
	\end{eqnarray}
and thus the degeneracy condition implies a single algebraic equation:
	\begin{equation}
	\mathcal{D}_{0}+\mathcal{D}_{1}X+\mathcal{D}_{2}X^{2} = 0. \label{degen_calD}
	\end{equation}

For the consistency condition, Eq. (\ref{calF_def}) can be written as
	\begin{eqnarray}
	\mathcal{F}\left(\vec{x},\vec{y}\right) & = & \partial_{y^{i}}\delta^{3}\left(\vec{x}-\vec{y}\right)\sqrt{h\left(\vec{y}\right)}\frac{1}{\Xi\left(\vec{y}\right)}\nonumber \\
	&  & \times\left[\left(\mathcal{E}_{0}+\mathcal{E}_{1}F+\mathcal{E}_{2}K+\mathcal{E}_{3}X_{kl}K^{kl}\right)\partial^{i}N+\mathcal{E}_{4}K^{ij}\partial_{j}N\right]\left(\vec{y}\right)\nonumber \\
	&  & -\left(\vec{x}\leftrightarrow\vec{y}\right), \label{con_consi}
	\end{eqnarray}
where
	\begin{eqnarray}
	\mathcal{E}_{0} & = & \frac{\partial a_{1}}{\partial X}f_{1}+\frac{\partial a_{2}}{\partial X}f_{2}+\left(a_{3}+\frac{\partial a_{3}}{\partial X}X\right)f_{3},\\
	\mathcal{E}_{1} & = & \frac{\partial c_{1}}{\partial X}f_{1}+2\frac{\partial c_{2}}{\partial X}f_{2}+\left(c_{3}+\frac{\partial c_{3}}{\partial X}X\right)f_{3},\\
	\mathcal{E}_{2} & = & \frac{\partial b_{1}}{\partial X}\left(f_{1}-f_{3}\right)+2\frac{\partial b_{2}}{\partial X}f_{1}+\left(b_{3}+\frac{\partial b_{3}}{\partial X}X\right)f_{3}+\frac{\partial c_{1}}{\partial X}f_{2},\\
	\mathcal{E}_{3} & = & \left(\frac{\partial b_{1}}{\partial X}+\frac{1}{2}b_{4}\right)f_{4}+\frac{\partial b_{3}}{\partial X}f_{1}+2\frac{\partial b_{4}}{\partial X}f_{3}+2\left(b_{5}+\frac{\partial b_{5}}{\partial X}X\right)f_{3}+\frac{\partial c_{3}}{\partial X}f_{2},\\
	\mathcal{E}_{4} & = & -\left(b_{1}+\frac{1}{2}b_{4}X\right)f_{4},
	\end{eqnarray}
with
	\begin{eqnarray}
	f_{1} & = & \left(b_{3}c_{3}-2b_{5}c_{1}\right)X^{2}+\left(-2b_{4}c_{1}-b_{1}c_{3}\right)X-3b_{1}c_{1},\\
	f_{2} & = & \left[2\left(b_{1}+2b_{2}\right)b_{5}-b_{3}^{2}\right]X^{2}+\left[4b_{2}b_{4}+2b_{1}\left(b_{3}+b_{4}\right)\right]X+2b_{1}\left(b_{1}+3b_{2}\right),\\
	f_{3} & = & \left[b_{3}c_{1}-\left(b_{1}+2b_{2}\right)c_{3}\right]X-b_{1}c_{1},\\
	f_{4} & = & \left(2b_{5}c_{1}-b_{3}c_{3}\right)X+3b_{3}c_{1}+2b_{4}c_{1}-2\left(b_{1}+3b_{2}\right)c_{3}.
	\end{eqnarray}
The consistency condition (\ref{consis_con}) thus implies
	\begin{equation}
	\mathcal{E}_{0}=\mathcal{E}_{1}=\mathcal{E}_{2}=\mathcal{E}_{3}=\mathcal{E}_{4}=0, \label{cons_calE}
	\end{equation}
which are differential equations for various coefficients.

\subsubsection{Coefficients quadratic in $X$}

Instead of solving (\ref{degen_calD}) and (\ref{cons_calE}) for coefficients $a_{1}$, etc., with general functional dependence on $X$,  in the following we assume all the coefficients $a_{1}$, etc., are polynomials of $X$ such that the Lagrangian in (\ref{S_quad}) involves up to the quadratic power of $X_{ij}$.
For simplicity, we assume 
	\begin{equation}
	a_{1}=a_{2}=a_{3}=0,
	\end{equation}
since the corresponding terms are linear in time derivatives and thus are irrelevant to the degeneracy of the kinetic terms.
According to the above assumption, we make the ansatz
	\begin{eqnarray}
	b_{1} & = & b_{10}+b_{11}X+b_{12}X^{2}, \label{b_1_form}\\
	b_{2} & = & b_{20}+b_{21}X+b_{22}X^{2},\\
	b_{3} & = & b_{30}+b_{31}X,\\
	b_{4} & = & b_{40}+b_{41}X,\\
	b_{5} & = & b_{50}, \label{b_5_form}
	\end{eqnarray}
and
	\begin{eqnarray}
	c_{1} & = & c_{10}+c_{11}X+c_{12}X^{2},\\
	c_{2} & = & c_{20}+c_{21}X+c_{22}X^{2},\\
	c_{3} & = & c_{30}+c_{31}X,
	\end{eqnarray}
where $b_{ij}$ and $c_{ij}$ have no $X$ dependence; i.e., they are functions of $t$ and $N$ only. In total, there are
19 coefficients. 
With this ansatz, the degeneracy and consistency conditions (\ref{degen_calD}) and (\ref{cons_calE}) yield a unique set of solutions:
	\begin{eqnarray}
	c_{11} & = & \frac{\left(2b_{11}+6b_{21}+b_{30}\right)c_{10}}{2\left(b_{10}+3b_{20}\right)},\\
	c_{12} & = & \frac{\left(2b_{12}+6b_{22}+b_{31}\right)c_{10}}{2\left(b_{10}+3b_{20}\right)},\\
	c_{20} & = & \frac{3c_{10}^{2}}{4\left(b_{10}+3b_{20}\right)},\\
	c_{21} & = & \frac{\left(3b_{11}+9b_{21}+3b_{30}+b_{40}\right)c_{10}^{2}}{4\left(b_{10}+3b_{20}\right)^{2}},\\
	c_{22} & = & \frac{\left(3b_{12}+9b_{22}+3b_{31}+b_{41}+b_{50}\right)c_{10}^{2}}{4\left(b_{10}+3b_{20}\right)^{2}},\\
	c_{30} & = & \frac{3b_{30}c_{10}+2b_{40}c_{10}}{2\left(b_{10}+3b_{20}\right)},\\
	c_{31} & = & \frac{\left[3b_{31}+2\left(b_{41}+b_{50}\right)\right]c_{10}}{2\left(b_{10}+3b_{20}\right)}.
	\end{eqnarray}
From the solutions it is clear that all $b_{ij}$'s are completely free, while among eight $c_{ij}$'s, only one (chosen as $c_{10}$) is independent, and another seven $c_{ij}$'s are determined by $b_{ij}$'s and $c_{10}$.

We introduce
	\begin{equation}
	\gamma\equiv\frac{3c_{10}}{2\left(b_{10}+3b_{20}\right)},
	\end{equation}
and
	\begin{eqnarray}
	b_{10}+3b_{20} & \equiv & 3\beta_{1},\\
	3b_{11}+9b_{21}+3b_{30}+b_{40} & \equiv & 9\beta_{2},\\
	3b_{12}+9b_{22}+3b_{31}+b_{41}+b_{50} & \equiv & 9\beta_{3},\\
	3b_{30}+2b_{40} & \equiv & 3\beta_{4},\\
	3b_{31}+2b_{41}+2b_{50} & \equiv & 3\beta_{5},
	\end{eqnarray}
which can be viewed as the generalization of (\ref{gamma_beta_1}) and (\ref{gamma_beta_2}).
With these new notations, we have
\begin{eqnarray}
c_{11} & = & \frac{1}{3}\gamma\left(6\beta_{2}-\beta_{4}\right),\\
c_{12} & = & \frac{1}{3}\gamma\left(6\beta_{3}-\beta_{5}\right),\\
c_{20} & = & \gamma^{2}\beta_{1},\\
c_{21} & = & \gamma^{2}\beta_{2},\\
c_{22} & = & \gamma^{2}\beta_{3},\\
c_{30} & = & \gamma\beta_{4},\\
c_{31} & = & \gamma\beta_{5}.
\end{eqnarray}

Finally, in terms of $\gamma$ and $\beta_{i}$'s, the healthy Lagrangian can be recast to be 
\begin{eqnarray}
\mathcal{L}^{\mathrm{(quad)}} & = & b_{1}\,\hat{K}_{ij}\hat{K}^{ij}+\hat{b}_{2}\left(K+\gamma F\right)^{2}+\hat{b}_{3}\,X^{ij}\hat{K}_{ij}\left(K+\gamma F\right)\nonumber \\
&  & +b_{4}\,X^{ij}\hat{K}_{ik}\hat{K}_{\phantom{k}j}^{k}+b_{5}\left(\hat{K}_{ij}X^{ij}\right)^{2},\label{L_quad_fin}
\end{eqnarray}
where $\hat{K}_{ij}$ is the traceless part of $K_{ij}$:
\begin{equation}
\hat{K}_{ij}:=K_{ij}-\frac{1}{3}K h_{ij}.
\end{equation}
Various coefficients are
\begin{eqnarray}
b_{1} & = & b_{10}+b_{11}X+b_{12}X^{2},\\
\hat{b}_{2} & = & \beta_{1}+\beta_{2}X+\beta_{3}X^{2},\\
\hat{b}_{3} & = & \beta_{4}+\beta_{5}X,\\
b_{4} & = & b_{40}+b_{41}X,\\
b_{5} & = & b_{50},
\end{eqnarray}
where keep in mind that $b_{ij}$'s and $\beta_{i}$'s contain no derivative of $N$.
Equation (\ref{L_quad_fin}) can be viewed as the generalization of (\ref{S_ori_fin}).

\subsection{Field transformations} \label{sec:trans}

We observe that $F \equiv \pounds_{\bm{n}} N$ enters (\ref{L_quad_fin}) in a specific manner, i.e., in terms of the special combination $K+\gamma F$.
In particular, terms with $F$ are controlled by a single coefficient $\gamma$. If we turn off $\gamma$, the Lagrangian (\ref{L_quad_fin}) will reduce to a special case of the spatially covariant gravity in Refs. \cite{Gao:2014soa,Gao:2014fra}.
This fact indicates that the Lagrangian (\ref{L_quad_fin}) may have some relation with theories in Refs. \cite{Gao:2014soa,Gao:2014fra}.

Let us consider the transformation of variables
\begin{eqnarray}
h_{ij} & \rightarrow & e^{2\omega}h_{ij}, \label{trans_h_ij}\\
N & \rightarrow & e^{\lambda}N, \label{trans_N}\\
N^{i} & \rightarrow & N^{i}, \label{trans_N^i}
\end{eqnarray}
where $\omega=\omega\left(t,N\right)$ and $\lambda=\lambda\left(t,N\right)$.
Generally we assume $\omega \neq \lambda$. The transformations (\ref{trans_h_ij})-(\ref{trans_N^i}) are nothing but correspond to the so-called disformal transformation of the metric \cite{Bekenstein:1992pj}. See also Appendix \ref{app:trans} for a brief discussion.

By definition, under the transformations (\ref{trans_h_ij})-(\ref{trans_N^i}), $X_{ij}$ transforms as
\begin{equation}
X_{ij}\rightarrow e^{2\lambda}\left(1+N\frac{\partial\lambda}{\partial N}\right)^{2}X_{ij},
\end{equation}
and it is easy to verify that
\begin{eqnarray}
K_{ij}  & \rightarrow & e^{2\omega-\lambda}\left(K_{ij}+h_{ij}\pounds_{\bm{n}}\omega\right)\nonumber \\
& = & e^{2\omega-\lambda}\left(K_{ij}+h_{ij}\frac{\partial\omega}{\partial N}\pounds_{\bm{n}}N\right).
\end{eqnarray}
Thus, generally, $F\equiv \pounds_{\bm{n}}N$ will arise after the transformation.
However, it is interesting that the traceless part of $K_{ij}$ transforms as
\begin{equation}
\hat{K}_{ij}\rightarrow e^{2\omega-\lambda}\hat{K}_{ij},
\end{equation}
in which $F$ drops out. On the other hand, the trace of $K_{ij}$ transforms as
	\begin{equation}
	K\rightarrow e^{-\lambda}\left(K+3\frac{\partial\omega}{\partial N}\pounds_{\bm{n}}N\right).
	\end{equation}
Comparing with (\ref{L_quad_fin}), it is thus clear that under the transformation (\ref{trans_h_ij})-(\ref{trans_N^i}), the Lagrangian
	\begin{eqnarray}
	\mathcal{L}^{\mathrm{(ori)}} & = & b_{1}\hat{K}_{ij}\hat{K}^{ij}+b_{2}K^{2}+b_{3}X^{ij}\hat{K}_{ij}K\nonumber \\
	&  & +b_{4}X^{ij}\hat{K}_{ik}\hat{K}_{\phantom{k}j}^{k}+b_{5}\left(\hat{K}_{ij}X^{ij}\right)^{2},
	\end{eqnarray}
which is a special case of the spatially covariant gravity proposed in Refs. \cite{Gao:2014soa,Gao:2014fra}, with coefficients taking the form (\ref{b_1_form})-(\ref{b_5_form}), is transformed to be exactly the form of (\ref{L_quad_fin}),
	\begin{eqnarray}
	\mathcal{L}^{\mathrm{(ori)}}\rightarrow\mathcal{L}^{\mathrm{(quad)}} & \equiv & \tilde{b}_{1}\hat{K}_{ij}\hat{K}^{ij}+\tilde{b}_{2}\left(K+\gamma F\right)^{2}\nonumber \\
	&  & +\tilde{b}_{3}X^{ij}\hat{K}_{ij}\left(K+\gamma F\right)\nonumber \\
	&  & +\tilde{b}_{4}X^{ij}\hat{K}_{ik}\hat{K}_{\phantom{k}j}^{k}+\tilde{b}_{5}\left(\hat{K}_{ij}X^{ij}\right)^{2},
	\end{eqnarray}
with
	\begin{equation}
	\gamma=3\frac{\partial\omega}{\partial N}
	\end{equation}
or, equivalently,
	\begin{equation}
	\omega\left(N\right)=\frac{1}{3}\int^{N}\mathrm{d}N'\,\gamma\left(N'\right).
	\end{equation}
The transformed coefficients are given by
\begin{eqnarray}
\tilde{b}_{1} & = & e^{-2\lambda}\left(b_{10}(\tilde{N})+b_{11}(\tilde{N})\tilde{X}+b_{12}(\tilde{N})\tilde{X}^{2}\right),\\
\tilde{b}_{2} & = & e^{-2\lambda}\left(b_{20}(\tilde{N})+b_{21}(\tilde{N})\tilde{X}+b_{22}(\tilde{N})\tilde{X}^{2}\right),\\
\tilde{b}_{3} & = & e^{-2\omega}\left(1+N\frac{\partial\lambda}{\partial N}\right)^{2}\left(b_{30}(\tilde{N})+b_{31}(\tilde{N})\tilde{X}\right),\\
\tilde{b}_{4} & = & e^{-2\omega}\left(1+N\frac{\partial\lambda}{\partial N}\right)^{2}\left(b_{40}(\tilde{N})+b_{41}(\tilde{N})\tilde{X}\right),\\
\tilde{b}_{5} & = & e^{-4\omega+2\lambda}\left(1+N\frac{\partial\lambda}{\partial N}\right)^{4}b_{50}(\tilde{N}),
\end{eqnarray}
with $\tilde{N} = e^{\lambda}N$, and
\begin{equation}
\tilde{X}=e^{-2\omega+2\lambda}\left(1+N\frac{\partial\lambda}{\partial N}\right)^{2}X.
\end{equation}
We thus conclude that the specific Lagrangian (\ref{L_quad_fin}), which depends on the $F$ term explicitly and is an explicit example of the formalism developed in Ref. \cite{Gao:2018znj}, can be related to a subclass of theories without $F$ term in Refs. \cite{Gao:2014soa,Gao:2014fra}, through a disformal transformation of the metric.

\section{Conclusion} \label{sec:con}

Recently, a general framework of spatially covariant theories of gravity was proposed in Ref. \cite{Gao:2018znj}, which generalized the theories in Refs. \cite{Gao:2014soa,Gao:2014fra} by including the kinetic terms of both the spatial metric $h_{ij}$ and the lapse function $N$. Generally, such a kind of theories propagates two scalar-type degrees of freedom.
Through a Hamiltonian analysis, a degeneracy condition and a consistency condition were found in order to evade the unwanted extra scalar mode.
In this work, we provide an alternative approach to and also a complementary understanding of such two conditions, by investigating the theory in a perturbative manner at the level of the Lagrangian.

First, we investigated the linear perturbations around a FRW background and showed that, as long as the kinetic terms for $h_{ij}$ and $N$ are degenerate, i.e., when the degeneracy condition (\ref{degen}) is satisfied, there is a single scalar mode that propagates. 
However, if only the degeneracy condition is imposed, the unwanted mode will generally reappear, either at the second order in perturbation around a FRW background or at the linear order in perturbation around an inhomogeneous background.
In both cases, we have shown that the same conditions (\ref{cons_sm_1a}) and (\ref{cons_sm_2a}), which correspond exactly to the consistency condition must be imposed in order to evade the unwanted mode.
According to such a perturbative analysis, it is interesting that to require the healthiness at the lowest a couple of orders in the perturbation theory (e.g., linear and second orders around the FRW background, or simply linear order around an inhomogeneous background) is enough to ensure the healthiness of the theory at all orders (or in a nonperturbative manner). 

In the final part of this work, we constructed another explicit example of the theories in Ref. \cite{Gao:2018znj}, which is quadratic in velocities of $h_{ij}$ and $N$, and up to the quadratic power in $X = \partial_{i}N \partial^{i}N$. It is interesting that the final Lagrangian with $\dot{N}$ can be generated by a disformal transformation of the metric from a Lagrangian without $\dot{N}$. In light of this result, one may study more general cases (e.g., higher-order powers in $K_{ij}$ and $F$ or with mixed spatial and time derivatives) to see if there exist healthy theories with $\dot{N}$ but cannot be transformed from theories without $\dot{N}$, which are genuinely new in the sense of field transformations.


\acknowledgments

This work was supported by the Chinese National Youth Thousand Talents Program (Grant No. 71000-41180003) and by the SYSU start-up funding.

\appendix

\section{Coefficients in $L_{2}^{(2)}(\zeta,A,B)$} \label{app:coeff} 

After making use of the background equations of motion (\ref{bgEoM_A}) and (\ref{bgEoM_z}), various coefficients in (\ref{L2^2}) are
	\begin{equation}
	\mathcal{C}_{\zeta^{2}}=-2d_{2}\frac{\partial^{2}}{a^{2}},
	\end{equation}
	\begin{equation}
	\mathcal{C}_{\dot{\zeta}A}=-3H\left[2\left(b_{1}+3b_{2}\right)-2\frac{\partial\left(b_{1}+3b_{2}\right)}{\partial N}+3c_{1}\right]+3\left(\frac{\partial a_{1}}{\partial N}-a_{2}\right),
	\end{equation}
	\begin{equation}
	\mathcal{C}_{\zeta A} = -4\left(d_{2}+\frac{\partial d_{2}}{\partial N}\right)\frac{\partial^{2}}{a^{2}},
	\end{equation}
	\begin{eqnarray}
	\mathcal{C}_{A^{2}} & = & -\frac{3}{2}\frac{\partial c_{1}}{\partial N}\dot{H}+\frac{1}{2}\left(d_{1}+3\frac{\partial d_{1}}{\partial N}+\frac{\partial^{2}d_{1}}{\partial N^{2}}\right)\nonumber \\
	&  & +\frac{3}{2}H^{2}\left[b_{1}+3b_{2}-\frac{\partial\left(b_{1}+3b_{2}\right)}{\partial N}+\frac{\partial^{2}\left(b_{1}+3b_{2}\right)}{\partial N^{2}}-3\frac{\partial c_{1}}{\partial N}\right]\nonumber \\
	&  & +\frac{3}{2}H\left(\frac{\partial a_{1}}{\partial N}-a_{2}+\frac{\partial^{2}a_{1}}{\partial N^{2}}-\frac{\partial a_{2}}{\partial N}\right)\nonumber \\
	&  & -\frac{1}{2}\left[3H\frac{\partial a_{1}}{\partial X}+3\frac{\partial\left(b_{1}+3b_{2}\right)}{\partial X}H^{2}+\frac{\partial d_{1}}{\partial X}\right]\frac{\partial^{2}}{a^{2}},
	\end{eqnarray}
	\begin{equation}
	\mathcal{C}_{AB}=\frac{1}{3}\mathcal{C}_{\dot{\zeta}A}.
	\end{equation}
In the above, terms involving $\partial^2$ are formal notations that must act on perturbation variables, which may be better understood in the Fourier space.

\section{Effective action for $\tilde{\zeta}$} \label{app:effact}

The coefficients in (\ref{S2_zetah_A}) are
\begin{eqnarray}
\mathcal{F}_{\dot{\zeta}A} & = & \mathcal{D}_{\dot{\zeta}A}-2\mathcal{D}_{\dot{\zeta}^{2}}\partial_{t}\sqrt{\frac{\mathcal{D}_{\dot{A}^{2}}}{\mathcal{D}_{\dot{\zeta}^{2}}}},\\
\mathcal{F}_{\zeta A} & = & \mathcal{C}_{\ensuremath{\zeta}A}-2\mathcal{C}_{\zeta^{2}}\sqrt{\frac{\mathcal{D}_{\dot{A}^{2}}}{\mathcal{D}_{\dot{\zeta}^{2}}}},\\
\mathcal{F}_{A^{2}} & = & \mathcal{D}_{A^{2}}+\mathcal{C}_{\zeta^{2}}\frac{\mathcal{D}_{\dot{A}^{2}}}{\mathcal{D}_{\dot{\zeta}^{2}}}+\mathcal{D}_{\dot{\zeta}^{2}}\left(\partial_{t}\sqrt{\frac{\mathcal{D}_{\dot{A}^{2}}}{\mathcal{D}_{\dot{\zeta}^{2}}}}\right)^{2}-\mathcal{C}_{\ensuremath{\zeta}A}\sqrt{\frac{\mathcal{D}_{\dot{A}^{2}}}{\mathcal{D}_{\dot{\zeta}^{2}}}}\nonumber \\
&  & -\mathcal{D}_{\dot{\zeta}A}\partial_{t}\sqrt{\frac{\mathcal{D}_{\dot{A}^{2}}}{\mathcal{D}_{\dot{\zeta}^{2}}}}+\frac{1}{2}\frac{1}{a^{3}}\partial_{t}\left(a^{3}\mathcal{D}_{\dot{\zeta}A}\sqrt{\frac{\mathcal{D}_{\dot{A}^{2}}}{\mathcal{D}_{\dot{\zeta}^{2}}}}\right).
\end{eqnarray}
From (\ref{S2_zetah_A}), the equation of motion for $A$ is
\begin{equation}
\mathcal{F}_{\dot{\zeta}A}\dot{\tilde{\zeta}}+\mathcal{F}_{\zeta A}\tilde{\zeta}+2\mathcal{F}_{A^{2}}A=0,
\end{equation}
from which we formally solve
\begin{equation}
A=-\frac{\mathcal{F}_{\dot{\zeta}A}}{2\mathcal{F}_{A^{2}}}\dot{\tilde{\zeta}}-\frac{\mathcal{F}_{\zeta A}}{2\mathcal{F}_{A^{2}}}\tilde{\zeta}.\label{A_sol}
\end{equation}
Finally, plugging (\ref{A_sol}) into (\ref{S2_zetah_A}) yields
	\begin{equation}
	S_{2}^{\mathrm{S}}[\tilde{\zeta}]=\int\mathrm{d}t\frac{\mathrm{d}^{3}k}{\left(2\pi\right)^{3}}a^{3}\left(\mathcal{G}\dot{\tilde{\zeta}}^{2}+\mathcal{W}\tilde{\zeta}^{2}\right),
	\end{equation}
with
\begin{eqnarray}
\mathcal{G} & \equiv & \mathcal{D}_{\dot{\zeta}^{2}}-\frac{\mathcal{F}_{\dot{\zeta}A}^{2}}{4\mathcal{F}_{A^{2}}},\\
\mathcal{W} & \equiv & \mathcal{C}_{\zeta^{2}}-\frac{\mathcal{F}_{\zeta A}^{2}}{4\mathcal{F}_{A^{2}}}+\frac{1}{4}\frac{1}{a^{3}}\partial_{t}\left(a^{3}\frac{\mathcal{F}_{\zeta A}\mathcal{F}_{\dot{\zeta}A}}{\mathcal{F}_{A^{2}}}\right).
\end{eqnarray}

\section{Cubic Lagrangian} \label{app:L3}

The cubic Lagrangian in (\ref{S3_zetaAB}) is
	\begin{eqnarray}
	\mathcal{L}_{3}^{\mathrm{S}} & = & A^{3}\frac{a^{3}}{6}\left(6\frac{\partial d_{1}}{\partial N}+6\frac{\partial^{2}d_{1}}{\partial N^{2}}+9\frac{\partial^{2}a_{1}}{\partial N^{2}}H+\mathcal{F}_{2}+\frac{\partial^{3}\mathcal{F}_{3}}{\partial N^{3}}\right)+A^{2}\zeta\,\frac{3}{2}a^{3}\mathcal{F}_{1}+A\zeta^{2}\,\frac{9}{2}a^{3}\mathcal{F}_{2}+\zeta^{3}\,\frac{9}{2}a^{3}\mathcal{F}_{3}\nonumber \\
	&  & -A^{2}\partial^{2}B\,\frac{a^{2}}{6}\frac{\partial\mathcal{F}_{1}}{\partial H}-A\zeta\partial^{2}B\,\frac{a^{2}}{3}\frac{\partial\mathcal{F}_{2}}{\partial H}-\zeta^{2}\partial^{2}B\,\frac{a^{2}}{6}\frac{\partial\mathcal{F}_{3}}{\partial H}-A^{2}\partial^{2}\zeta\,2a\left(d_{2}+3\frac{\partial d_{2}}{\partial N}+\frac{\partial^{2}d_{2}}{\partial N^{2}}\right)\nonumber \\
	&  & -A\zeta\partial^{2}\zeta\,4a\left(d_{2}+\frac{\partial d_{2}}{\partial N}\right)-\zeta^{2}\partial^{2}\zeta\,2ad_{2}+A\partial_{i}A\partial^{i}A\,\frac{a}{2}\left(2\frac{\partial d_{1}}{\partial X}+3\frac{\partial a_{1}}{\partial X}H+\frac{\partial\mathcal{F}_{3}}{\partial X}+\frac{\partial^{2}\mathcal{F}_{3}}{\partial N\partial X}\right)\nonumber \\
	&  & +\zeta\partial_{i}A\partial^{i}A\,\frac{a}{2}\frac{\partial\mathcal{F}_{3}}{\partial X}-A\partial_{i}B\partial^{i}A\,a^{2}\left(a_{2}+\frac{\partial\mathcal{F}_{4}}{\partial N}\right)-\zeta\partial_{i}B\partial^{i}A\,a^{2}\mathcal{F}_{4}-A\partial_{i}\zeta\partial^{i}B\,\frac{a^{2}}{3}\frac{\partial\mathcal{F}_{2}}{\partial H}\nonumber \\
	&  & -\zeta\partial_{i}\zeta\partial^{i}B\,a^{2}\frac{1}{3}\frac{\partial\mathcal{F}_{3}}{\partial H}-A\partial_{i}\zeta\partial^{i}\zeta\,2a\left(d_{2}+\frac{\partial d_{2}}{\partial N}\right)-\zeta\partial_{i}\zeta\partial^{i}\zeta\,2ad_{2}-A\partial^{2}B\partial^{2}B\,a\left(b_{2}+\frac{\partial b_{2}}{\partial N}\right)\nonumber \\
	&  & -\zeta\partial^{2}B\partial^{2}B\,ab_{2}-\partial_{i}A\partial^{i}A\partial^{2}B\,\frac{1}{6}\frac{\partial^{2}\mathcal{F}_{3}}{\partial H\partial X}+\partial_{i}B\partial^{i}A\partial^{2}B\,ac_{1}+\partial_{i}\zeta\partial^{i}B\partial^{2}B\,2a(b_{1}+b_{2})\nonumber \\
	&  & -\partial_{i}A\partial^{i}A\partial^{2}\zeta\,\frac{2}{a}\frac{\partial d_{2}}{\partial X}-\partial^{i}B\partial_{j}\partial_{i}B\partial^{j}\zeta\,4ab_{1}-A\partial_{j}\partial_{i}B\partial^{j}\partial^{i}B\,a\left(b_{1}-\frac{\partial b_{1}}{\partial N}\right)-\zeta\partial_{j}\partial_{i}B\partial^{j}\partial^{i}B\,ab_{1}\nonumber \\
	&  & +A^{2}\dot{A}\,\frac{a^{3}}{2}\left(a_{2}+2\frac{\partial a_{2}}{\partial N}+\frac{\partial\mathcal{F}_{4}}{\partial N}+\frac{\partial^{2}\mathcal{F}_{4}}{\partial N^{2}}\right)+\zeta A\dot{A}\,3a^{3}\left(a_{2}+\frac{\partial\mathcal{F}_{4}}{\partial N}\right)+\zeta^{2}\dot{A}\,\frac{9}{2}a^{3}\mathcal{F}_{4}+A^{2}\dot{\zeta}\,\frac{a^{3}}{2}\frac{\partial\mathcal{F}_{1}}{\partial H}\nonumber \\
	&  & +A\zeta\dot{\zeta}\,3a^{3}\frac{\partial\mathcal{F}_{2}}{\partial H}+\zeta^{2}\dot{\zeta}\frac{9}{2}a^{3}\frac{\partial\mathcal{F}_{3}}{\partial H}+\partial_{i}A\partial^{i}A\dot{A}\,\frac{a}{2}\frac{\partial\mathcal{F}_{4}}{\partial X}+\partial_{i}A\partial^{i}A\dot{\zeta}\,\frac{a}{2}\frac{\partial^{2}\mathcal{F}_{3}}{\partial H\partial X}-A\dot{A}\partial^{2}B\,a^{2}\frac{\partial c_{1}}{\partial N}\nonumber \\
	&  & -\zeta\dot{A}\partial^{2}B\,a^{2}c_{1}-\partial_{i}B\partial^{i}A\dot{A}\,2a^{2}c_{2}-\partial_{i}\zeta\partial^{i}B\dot{A}\,a^{2}c_{1}+A\dot{A}^{2}a^{3}\left(c_{2}+\frac{\partial c_{2}}{\partial N}\right)+\zeta\dot{A}^{2}3a^{3}c_{2}\nonumber \\
	&  & -A\partial^{2}B\dot{\zeta}\,\frac{a^{2}}{3}\frac{\partial^{2}\mathcal{F}_{2}}{\partial H^{2}}-\partial^{2}B\zeta\dot{\zeta}\,2a^{2}\left(b_{1}+3b_{2}\right)-\partial^{i}A\partial_{i}B\dot{\zeta}\,3a^{2}c_{1}-\partial^{i}B\partial_{i}\zeta\dot{\zeta}\,2a^{2}(b_{1}+3b_{2})\nonumber \\
	&  & +A\dot{A}\dot{\zeta}\,3a^{3}\frac{\partial c_{1}}{\partial N}+\dot{A}\zeta\dot{\zeta}\,9a^{3}c_{1}+A\dot{\zeta}^{2}\frac{a^{3}}{2}\frac{\partial^{2}\mathcal{F}_{2}}{\partial H^{2}}+\zeta\dot{\zeta}^{2}\,9a^{3}\left(b_{1}+3b_{2}\right), \label{L3_zetaAB}
	\end{eqnarray}
where we introduce
	\begin{eqnarray}
	\mathcal{F}_{1} & \equiv & d_{1}+3\frac{\partial d_{1}}{\partial N}+\frac{\partial^{2}d_{1}}{\partial N^{2}}+3\left(\frac{\partial a_{1}}{\partial N}+\frac{\partial^{2}a_{1}}{\partial N^{2}}\right)H\nonumber \\
	&  & +3\left(b_{1}+3b_{2}-\frac{\partial b_{1}(b_{1}+3b_{2})}{\partial N}+\frac{\partial^{2}(b_{1}+3b_{2})}{\partial N^{2}}\right)H^{2},
	\end{eqnarray}
	\begin{eqnarray}
	\mathcal{F}_{2} & \equiv & d_{1}+\frac{\partial d_{1}}{\partial N}+3\frac{\partial a_{1}}{\partial N}H\nonumber \\
	&  & -3\left(b_{1}+3b_{2}-\frac{\partial b_{1}}{\partial N}-3\frac{\partial b_{2}}{\partial N}\right)H^{2},
	\end{eqnarray}
	\begin{equation}
	\mathcal{F}_{3}\equiv d_{1}+3a_{1}H+3(b_{1}+3b_{2})H^{2},
	\end{equation}
	\begin{equation}
	\mathcal{F}_{4}\equiv a_{2}+3c_{1}H,
	\end{equation}
for shorthand.
Please note that we have not performed any integration by parts to simply (\ref{L3_zetaAB}), which is not necessary for our purpose.

\section{Field transformation} \label{app:trans}

In four-dimensional formalism, let us consider the transformation of the induced metric and the normal vector of the spatial hypersurfaces,
\begin{eqnarray}
h_{\mu\nu}\rightarrow\tilde{h}_{\mu\nu} & \equiv & e^{2\omega}h_{\mu\nu},\label{h_munu_trans}\\
n^{\mu}\rightarrow\tilde{n}^{\mu} & \equiv & e^{-\lambda}n^{\mu},\label{n^mu_trans}
\end{eqnarray}
where $\omega,\lambda$ are general functions of $t,N$ (with $n_{\mu}\equiv-N\partial_{\mu}t$).
The normalization of $\tilde{n}^{\mu}$, i.e., $-1=\tilde{n}^{\mu}\tilde{n}_{\mu}=e^{-\lambda}n^{\mu}\tilde{n}_{\mu}$,
implies
\begin{equation}
n_{\mu}\rightarrow\tilde{n}_{\mu}=e^{\lambda}n_{\mu}.
\end{equation}
Thus we also have
	\begin{equation}
	N\rightarrow\tilde{N}=e^{\lambda}N,
	\end{equation}
and
	\begin{equation}
		N^{\mu} \rightarrow \tilde{N}^{\mu}=N^{\mu}.
	\end{equation}

The transformation (\ref{h_munu_trans})-(\ref{n^mu_trans}) also imply the  transformation for the metric $g_{\mu\nu}$
\begin{eqnarray}
g_{\mu\nu}\rightarrow\tilde{g}_{\mu\nu} & = & \tilde{h}_{\mu\nu}-\tilde{n}_{\mu}\tilde{n}_{\nu}\nonumber \\
& = & e^{2\omega}h_{\mu\nu}-e^{2\lambda}n_{\mu}n_{\nu}\nonumber \\
& = & e^{2\omega}\left(g_{\mu\nu}+n_{\mu}n_{\nu}\right)-e^{2\lambda}n_{\mu}n_{\nu}\nonumber \\
& = & e^{2\omega}g_{\mu\nu}+\left(e^{2\omega}-e^{2\lambda}\right)n_{\mu}n_{\nu},
\end{eqnarray}
which is a disformal transformation for $g_{\mu\nu}$ if $\omega \neq \lambda$.
The transformed inverse metric takes the form
	\begin{equation}
	\tilde{g}^{\mu\nu}=e^{-2\omega}g^{\mu\nu}+\left(e^{-2\omega}-e^{-2\lambda}\right)n^{\mu}n^{\nu}.
	\end{equation}
Using the above results, we also have
\begin{eqnarray}
	\tilde{h}^{\mu\nu} & = & e^{-2\omega}h^{\mu\nu},\\
	\tilde{h}_{\mu}^{\phantom{\mu}\nu} & = & h_{\mu}^{\phantom{\mu}\nu}.
\end{eqnarray}


%

\end{document}